\documentclass{cccg18}

\usepackage[utf8]{inputenc}
\usepackage[T1]{fontenc}
\usepackage{amsmath,amssymb}
\usepackage[ruled,linesnumbered]{algorithm2e}
\usepackage{pgfplots}
\usepackage{microtype}
\usepackage{xspace}
\usepackage{csquotes}
\usepackage[multiple]{footmisc}
\usepackage{listings}
\title{On error representation in exact-decisions number types}
\author{Martin Wilhelm\thanks{Institut für Simulation und Graphik,
        Otto-von-Guericke-Universität Magdeburg, {\tt martin.wilhelm@ovgu.de}}}

\newcommand{\RealAlgebraic}{\texttt{Real\_algebraic}\xspace}
\newcommand{\LedaReal}{\texttt{leda::real}\xspace}
\newcommand{\CoreExpr}{\texttt{Core::Expr}\xspace}
\newcommand{\cpp}{C\texttt{++}}
\newcommand{\gpp}{g\texttt{++}}

\newcommand{\deferr}{{e}}
\newcommand{\direrr}{{e_{dir}}}
\newcommand{\logerr}{{e_{log}}}
\newcommand{\rval}{x}
\newcommand{\apx}{\hat{x}}
\newcommand{\qmax}{q_{\max}}

\newcommand{\tlogdir}{\Phi}
\newcommand{\tdirlog}{\hat{\Phi}}

\DeclareMathOperator{\sep}{sep}
\DeclareMathOperator{\val}{value}

\definecolor{SEQCOL1}{RGB}{227,74,51}
\definecolor{SEQCOL2}{RGB}{253,187,132}
\definecolor{SEQCOL3}{RGB}{254,232,200}

\definecolor{QCOL1}{RGB}{241,163,64}
\definecolor{QCOL2}{RGB}{153,142,195}

\begin{document}
\maketitle

\begin{abstract}
Accuracy-driven computation is a strategy widely used in exact-decisions number types for robust geometric algorithms.
This work provides an overview on the usage of error bounds in accuracy-driven computation, compares different approaches on the representation and computation of these error bounds and points out some caveats. The stated claims are supported by experiments.
\end{abstract}

\section{Introduction}

In Computational Geometry, many algorithms rely on the correctness of geometric predicates, such as orientation tests or incircle tests. In contrast to various other areas of computing, a small error in computation often does not imply a small error in the result. Instead, algorithms may fail, produce drastically wrong output, or no output at all if a predicate returns the wrong result~\cite{schirra2000}.

To mitigate the consequences of this problem, exact-decisions number types have been developed, based on the concept of accuracy-driven-computation and the Exact Geometric Computation Paradigm~\cite{yap1997}. Examples for such number types are \LedaReal from the LEDA library~\cite{burnikel1996}, \CoreExpr~\cite{karamcheti1999,yu2010} and \RealAlgebraic~\cite{moerig10}. They store the expressions involved in directed acyclic graphs, called expression dags, and maintain approximations and error bounds for each subexpression. When a decision has to be made, the accuracy of the subexpressions is increased until the value can be separated from zero or value can be guaranteed to be zero through a separation bound~\cite{burnikel2009}.
This concept will be explained in slightly more detail in Section~\ref{ssc:accuracydriven}.

Approximations are usually stored in ar\-bi\-tra\-ry-pre\-ci\-sion floating-point number types, or short, \emph{bigfloats}. In both \LedaReal and \RealAlgebraic, error bounds are stored in a bigfloat as well, in form of an absolute error bound. In \CoreExpr, error bounds are stored as a combination of upper and lower bound for the most significant bit of the value.

In this paper we will show advantages and disadvantages of different error bound representation forms for accuracy-driven computation. In Section~\ref{sec:errorrepresentation}, various possibilities are evaluated and sensible choices for an implementation are proposed. We present an experimental comparison for these choices in Section~\ref{sec:experiments} based on the number type \RealAlgebraic.

\section{Error Representation}
\label{sec:errorrepresentation}

There are more aspects to the representation of an error bound than might be apparent at first glance. One natural way to represent an error is by storing a value $\direrr$, such that $\apx - \direrr \leq \rval \leq \apx + \direrr$, if $\rval$ represents the real value of the expression and $\apx$ its approximation. We call this the \emph{direct} error representation.

Since an error bound can get quite small, a direct representation must be stored in a non-primitive data type, most commonly in a bigfloat. Computations involving bigfloats are expensive. Another natural way of storing an error bound is to store an exponent $\logerr$, such that $\apx - 2^\logerr \leq \rval \leq \apx + 2^\logerr$. It then suffices to store $\logerr$ as a primitive integer data type, such as \texttt{long}. We call this the \emph{logarithmic} representation.

The obvious advantage of a direct over a logarithmic value is that the direct representation can be much more precise than the logarithmic one. If, for example, a large sum is computed with error $\deferr$ at the $m$ operands, error propagation with a direct representation leads to an error bound of $m\deferr$ (assuming exact operations). A logarithmic integer bound, on the other hand, increases by at least $1$ for each addition. So error propagation leads to an error bound of $2^{\log(\deferr)+m}$ if the additions are processed sequentially.

To keep the bound small while avoiding bigfloats, a third approach on error representation would be to store a logarithmic error bound in a floating-point primitive, such as \texttt{double}. While not as precise as the direct representation, the error bound can be increased in smaller steps, avoiding an exponential increase as in the previous example. In contrast to the error representation as an integer value, it might also enable more elaborate evaluation strategies, such as proposed by van der Hoeven~\cite{vanderHoeven2006}.

There are various other ways to represent an error. While all three methods above store the radius of an error interval, the interval can be stored directly through an upper and lower bound, such as in \CoreExpr. Furthermore one can imagine various combined approaches, where the representation can change based on the size of the error. In this paper, however, we will focus on the three variants proposed above. We also only consider absolute error bounds, although there are good reasons to compute the bound relative to the size of the approximation or even combine an absolute and a relative error~\cite{ouchi1997}.

\subsection{Errors in accuracy-driven computation}
\label{ssc:accuracydriven}

We introduced three different ways of representing an error bound in a number type. The complexity rises if we consider different combinations of the approaches during the computation process. The usefulness of error bound representations may change depending on the current task. So it could be advantageous to switch between different representations during the computation.

In this section we will shortly describe the concept of accuracy-driven computation as implemented in \RealAlgebraic and the usage of error bounds within. Accuracy-driven computation was first introduced by Yap and Dubé under the name ``precision-driven computation''~\cite{yap1995}. It describes a lazy approach on exact-decisions computation. Computations are not done directly on invocation, but stored in an \emph{arithmetic expression dag}, i.e., a rooted ordered directed acyclic graph whose nodes are either a floating-point number or an operator with its operands as children.

Once a decision has to be made, the computation is started. Since every decision can be translated to the decision, whether a value is positive, negative or zero, it is sufficient to determine the sign of the root node. This can be done by using a separation bound, i.e., a number $\sep(E)$ for an expression $E$, such that $|\val(E)| > 0 \Rightarrow |\val(E)| > \sep(E)$. If $\sep(E)$ is not part of the error interval of an approximation, then the sign of the approximation is correct.

The strategy in accuracy-driven computation is to start with a desired accuracy $q$ at a (root) node, compute the accuracy needed at its children to guarantee a respective error bound, and recurse on the children with the new desired accuracies. To determine whether an expression $E$ is zero, an accuracy of $\qmax = \lfloor\log(\sep(E))\rfloor$ is needed. If $\val(E) > 0$, however, a separation is usually possible with a much larger error as soon as zero is not contained anymore in the error interval. So usually $q$ is chosen to be a small negative number\footnote{We will refer to accuracies as ``small'' if their absolute value is small, although they are usually negative numbers. We fear that the opposite notion would be even more misleading.} in the beginning, with an exponential increase until $|q| > |\qmax|$.

Before the top-down computation is started, an initial value for each node must be computed bottom-up, i.e., with a small fixed precision. This is necessary, since sometimes an estimate for the value of a child node is needed to compute the required accuracy. During this process, an accuracy-driven-computation for a subexpression may be invoked, if the value of a divisor or the operand of a root needs to be separated from zero.

Error bounds mainly occur in three different places:
\begin{enumerate}
\item In each node a current error bound is stored to prevent recomputations of approximations that are already sufficiently accurate.
\item A desired error bound is computed and propagated top-down through the expression dag.
\item An initial error bound for each node is computed bottom-up.
\end{enumerate}

In \RealAlgebraic, a direct error bound is used in the first and the third case, while the top-down propagation is done with a logarithmic integer error bound. The number type \LedaReal uses direct error bounds in all three cases, whereas \CoreExpr uses a logarithmic integer bound in the second and a logarithmic integer interval, i.e., a combination of upper and lower bound in the third case. Both representations are saved inside each node. 

The initial bottom-up computation is done with small precision. Bigfloat operations are less expensive then, whereas the influence of a weak error bound is increased. A direct error bound is therefore a sensible choice for this part of the algorithm. In contrast, the main accuracy-driven parts of the computation require high precisions, causing the maintenance of a direct error bound to be too expensive compared to its benefits~\cite{moerig15}.

\subsection{Switching between direct and logarithmic bound}
\label{ssc:switchingerrors}

\setlength{\thickmuskip}{5mu plus 5mu minus 3mu}
When parts of the algorithm are computed with differing error representations, the algorithm must switch between those representations. Since floating-point data types are stored as a mantissa and an exponent, computing a direct bound $\direrr$ from a logarithmic bound $\logerr$ can be done fast and without loss of precision by setting the exponent of $\direrr$ to $\logerr$, i.e., $\direrr=\tlogdir(\logerr):=2^\logerr$. 
For the reverse process, $\logerr$ must be computed as $\logerr = \tdirlog(\direrr) := \lceil\log(\direrr)\rceil$, losing some precision in the progress. 
In particular, it cannot be expected that $\direrr = \tlogdir(\tdirlog(\direrr))$. However, we should expect $\logerr=\tdirlog(\tlogdir(\logerr))$ to be true.
\setlength{\thickmuskip}{5mu plus 5mu}

As previously mentioned, computing $\tlogdir$ is cheap. What about $\tdirlog$? The mantissa $m$ and exponent $b$ of a floating-point value $x$ are usually chosen, such that $m\in [0.5,1)$, $b\in\mathbb{Z}$ and $x = m2^b$. So it seems natural to choose $\tdirlog(x) = b$ as a cheap conversion function, as done in \RealAlgebraic.

However, there is a significant drawback to this approach. If $x$ is a power of two, $\tdirlog$ overestimates $\lceil\log(x)\rceil$ by one, since then $m=0.5$. While this does not affect overall correctness of the algorithm and the case seems very special, implementing $\tdirlog$ like that can lead to massive drops in performance, since then $\logerr \neq \tdirlog(\tlogdir(\logerr))$ for every value of $\logerr$.

After an error bound is guaranteed by the accuracy-driven computation, the stored error is set to this error bound. If later on the same error bound is needed for the respective node, the algorithm checks whether a recomputation is necessary. If the representations of the stored bound and the requested bound differ, as present in \RealAlgebraic, the check fails and the computation needs to be executed again.

\begin{figure}[ht]
\centering
\includegraphics[width=0.7\linewidth]{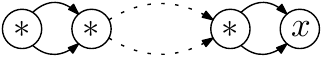}
\caption{The arithmetic expression dag resulting from computing $x^{2^n}$ through repeated squaring. There are $2^n$ different paths from the topmost multiplication to the common leaf $x$. If recomputation checks fail due to conversion errors, the evaluation time increases drastically.}
\label{fig:repeatedsquaring}
\end{figure}

A drastic example for this effect arises if a power is computed through repeated squaring. Each node gets recomputed along each of the $2^n$ paths from the root to the leaf, leading to an exponential increase in running time (cf.\ Figure~\ref{fig:repeatedsquaring}). For $x=\sqrt{13}+\sqrt{17}$ and $n=15$ operations, \RealAlgebraic takes $87.84$ seconds with the ``inexact'' implementation to evaluate the expression up to an accuracy of $q=50000$, compared to about $0.01$ seconds with an ``exact'' one\footnote{We use the term ``exact'' in this context to express that the method correctly implements the function $\tdirlog(\direrr) := \lceil\log(\direrr)\rceil$.}\footnote{Specifications on the test configuration can be found at the beginning of Section~\ref{sec:experiments}.}. While in this example the problem can be avoided by switching to a topological evaluation order (see Mörig et al.~\cite{moerig15macis}), it persists if checks need to be repeated in the main algorithm.

If the conversion should be exact, considerable effort needs to be done. We need to check, whether the value of the mantissa is exactly $0.5$ and, if so, decrease the result by one. When using \texttt{mpfr} bigfloats for example, this changes the one-liner

\begin{lstlisting}
mpfr_exp_t ceil_log2(const mpfr_t& a){
	return mpfr_get_exp(a);
}
\end{lstlisting}
to the more elaborate method
\begin{lstlisting}
mpfr_exp_t ceil_log2(const mpfr_t& a){
	mpfr_exp_t e = mpfr_get_exp(a);
	mpfr_t rop;	mpfr_init(rop);
	mpfr_div_2si(rop,a,e,MPFR_RNDA);
	if (mpfr_cmp_d(rop,0.5) == 0) --e;
	mpfr_clear(rop);
	return e;
}
\end{lstlisting}

In the second method, the mantissa of the bigfloat is accessed, which can potentially be large. For large computations it can be expected to be significantly slower than the first method. If the undesirable effects described above should be avoided, it may therefore prove more efficient to avoid transformations between different error representations entirely.

Note that even then the conversion function is widely used during the computation. Approximations of the value of a (sub)expression must be stored in bigfloats. In accuracy-driven computation, often a bound for the magnitude of the result is needed, which is then computed by the above function. Since, in contrast to error transformation, a worse bound for the magnitude only causes a small difference in performance, it is reasonable to use the inexact, but fast transformation method in those cases.

\subsection{Logarithmic floating-point error bounds}
\label{ssc:logdbl}

In the previous section, difficulties are pointed out that may arise when switching between a direct and a logarithmic error representation. In contrast, switching between a floating-point representation and an integer representation for logarithmic errors is cheap and natural. This raises the question whether the fixed precision computations can be done more efficiently through a logarithmic floating-point bound.

Most of the computations involved in error propagation with a fixed precision can be broken down to the sum of two or three errors. So with a logarithmic bound we have to find a value $c$ for two error representations $a$ and $b$, such that $2^c\geq 2^a + 2^b$. With an integer value the error bound doubles with each such summation, since this is the smallest step in which the bound can be increased. So we have $c=\max(a,b)+1$.

If $a,b,c$ are floating-point values, we may find a better bound by setting $c=\max(a,b)+\log(1+2^{-|a-b|})$, where we have to make sure that each operation is rounded up (towards infinity). With repeated additions, the floating-point error bound increases much more slowly than the integer bound, although computing the logarithm in each step makes it also a lot more expensive.

Can the error propagation during accuracy-driven computation benefit from more precision in the exponent? Surprisingly, the answer is no, at least not directly. When deciding which accuracy is needed at the child nodes to guarantee a certain accuracy at the parent node, up to three error terms need to be balanced. Besides the desired accuracies of the one or two children, the precision at which the operation at the parent node is computed must be chosen. A higher accuracy at the child nodes can then be used to reduce the precision needed for the bigfloat operation.

This decision is done locally, i.e., the parent node does neither know of what size, nor of which form its subtrees are. Without this information it cannot be decided to what extent the precision of the operation needs to be increased in order to require a smaller overall accuracy. So if the decision should be made locally, the gain from switching to a floating-point exponent is marginal and will probably not cover the additional costs associated with it.

Nevertheless, benefitting may be possible if a global error propagation strategy is implemented. The overall accuracy needed could then be kept small through balancing of error terms, which in turn has the potential to drastically improve the performance of the number type, especially for unbalanced expression dags~\cite{vanderHoeven2006,wilhelm17}.

\subsection{Errors and separation bounds}
\label{ssc:errorsepbounds}

During the accuracy-driven part of the computation, it is beneficial to convert subgraphs to a single bigfloat node if their approximation is already exact, i.e., if their error is zero. This is especially useful if the value of a subgraph is found to be zero. If after a computation the approximation of a node is close to zero, \RealAlgebraic computes a separation bound for this node and checks whether the true value can be declared zero. An approximation is considered close to zero if zero is part of its error interval.

\begin{figure}[ht]
\centering
\includegraphics[scale=2]{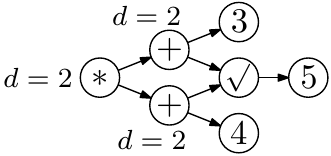}
\caption{The two children of the root node share a common subexpression with a square root operation. Although both subexpression at the child nodes have algebraic degree two, the algebraic degree of the full expression is still only two.}
\label{fig:algdegree}
\end{figure}

If the error bound is bad, this check happens (and fails) more often. This can have significant consequences for the performance, since for computing the separation bound of a subexpression, the whole subtree must be traversed. Existing bounds for the child nodes cannot be used, since they may misrepresent the algebraic degree of the expression if common subexpressions exist (cf.\ Figure~\ref{fig:algdegree}) and a higher algebraic degree drastically worsens the separation bound.

In addition, the separation bound cannot be assumed to stay the same during the whole computation, since subgraphs may be converted to bigfloat nodes. While a previous separation bound stays valid after such a conversion, in some cases a much better bound can be computed, e.g.\ if roots can be eliminated. So it is advisable to keep the bound flexible.

To solve this problem, once computed separation bounds can be cached together with a global timestamp. Whenever a bigfloat conversion happens, all previous timestamps get invalidated. The advantage of this method is an easy implementation without much overhead, leading to good results, if the usage of the number type is limited to few large expressions. However, if many different usages of the number type exist at the same time, a separation bound might get invalidated by a bigfloat conversion in a completely disjoint expression dag. An alternative approach based on topological evaluation which leads to similar results can be found in~\cite{wilhelm18techreport}.

\section{Experimental Results}
\label{sec:experiments}

The experiments are performed on an Intel Core i5 660, 8GB RAM, under Ubuntu 17.10. For \RealAlgebraic we use Boost interval arithmetic as floating-point-filter and MPFR bigfloats for the bigfloat arithmetic. The code is compiled  using \gpp~7.2.0 with \cpp11 on optimization level \texttt{O3} and linked against Boost~1.62.0 and MPFR~3.1.0. Test results are averaged over 25 runs each. The variance for each data point is negligible.

We compare three main strategies with different representations for (a) storing, (b) error propagation in accuracy-driven computation and (c) fixed precision evaluation (cf. Section~\ref{ssc:accuracydriven}):
\begin{enumerate}
	\item The default strategy of \RealAlgebraic, i.e., direct error representation for (a) and (c) and logarithmic integer representation for (b), named \texttt{def}.
	\item Logarithmic integer representation for all three parts (a), (b) and (c), named \texttt{lgi}.
	\item Logarithmic floating-point representation for (a) and (c), logarithmic integer representation for (b), named \texttt{lgd}.
\end{enumerate}

Note, that in each case we use a logarithmic integer representation for (b). It has already been shown that direct error bounds are very expensive compared to the additional benefit~\cite{moerig15}. As described in Section~\ref{ssc:logdbl}, advanced strategies would be needed for error propagation to benefit from a logarithmic floating-point representation. While without such strategies, no interesting results are to be expected, implementing them heavily reduces comparability to the other approaches presented in this paper. Therefore we leave it aside for future work.

\subsection{Comparison of separation bound strategies}

First, we show the effects of the interactions between error bound representation and the computation of separation bounds, as described in Section~\ref{ssc:errorsepbounds}.
We test for the equality \[F_n = \frac{\phi^n-\bar{\phi}^n}{\sqrt{5}\phantom{1}},\] where $F_n$ represents the $n$-th Fibonacci number and $\phi = 1-\bar{\phi} = \frac{1+\sqrt{5}}{2}$. We compute both sides of the equation in a simple loop as in the code below.

\begin{lstlisting}
template <class NT>
bool fibonacci_test(const int n){
	NT sqrt5  = sqrt(NT(5));
	NT phi    = (NT(1) + sqrt5) / NT(2);
	NT phibar = (NT(1) - sqrt5) / NT(2);

	NT phiN = phi; NT phibarN = phibar;
	NT fib0 = 0; NT fib1 = 1; NT tmp;

	for(int i = 1; i < n; i++){
		tmp = fib1; fib1 += fib0; fib0 = tmp;
		phiN *= phi; phibarN *= phibar;
	}
	
	NT res = NT(1)/sqrt5 * (phiN-phibarN);
	return fib1 == res;
}
\end{lstlisting}

We show running times for each of the three aforementioned strategies with and without a cached separation bound computation (indicated by an additional \texttt{s}).

\begin{figure}[ht]
\begin{tikzpicture}[scale=0.5]
\begin{axis}[
	symbolic x coords={def,lgi,lgd,defs,lgis,lgds},
	xtick=data,
	ylabel=Time (seconds),
	ymax = 17,
    width  = 1.9\linewidth,
    height = \linewidth,
    bar width = 0.7cm,
    ybar,
    label style={font=\Large},
    tick label style={font=\Large},
    legend style={font=\Large},
    nodes near coords,
    every node near coord/.append style={rotate=90, anchor=west, 
    									 /pgf/number format/precision=4,font=\Large}
]
%50k

\addplot[fill=SEQCOL1] coordinates {(def,0.25) (lgi,0.27) (lgd,0.25) (defs,0.25) (lgis,0.2) (lgds,0.22) };
\addplot[fill=SEQCOL2] coordinates {(def,1.47) (lgi,2.06) (lgd,1.7) (defs,1.47) (lgis,0.84) (lgds,1.06) };
\addplot[fill=SEQCOL3] coordinates {(def,8.87) (lgi,12.38) (lgd,10.23) (defs,8.42) (lgis,3.91) (lgds,5.41) };

\legend{$n=2000$,$n=4000$,$n=8000$}
\end{axis}
\end{tikzpicture}
\caption{Results for \texttt{fibonacci\_test} with different strategies. A logarithmic error bound increases the running time due to repeated computation of separation bounds. If the separation bounds are cached, the running time can be significantly decreased.}
\label{fig:sepbd}
\end{figure}
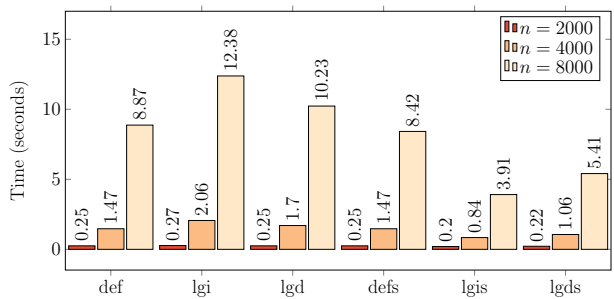

The results are shown in Figure~\ref{fig:sepbd}. Representing the error logarithmically has a negative impact on the performance due to a heavy increase in separation bound computations. For $n=8000$, a separation bound computation is started $1149$ times with \texttt{def} compared to $29092$ and $13939$ times with \texttt{lgi} and \texttt{lgd}. When the cost for those computations can be reduced due to caching, the logarithmic error bounds outperform the direct representation. Note that the logarithmic floating-point representation ranges in between the other two representations in each scenario, which underlines its character of slightly more precision for slightly more overhead compared to a logarithmic integer representation.

\subsection{Geometric problems}

In the previous section a first impression of the behavior of the three different representations is obtained. In this section we test this impression against more realistic geometric problems. For this, we recreate several experiments from Mörig et al.~\cite{moerig10}.

\begin{figure}[ht]
\centering
\includegraphics[width=0.32\linewidth]{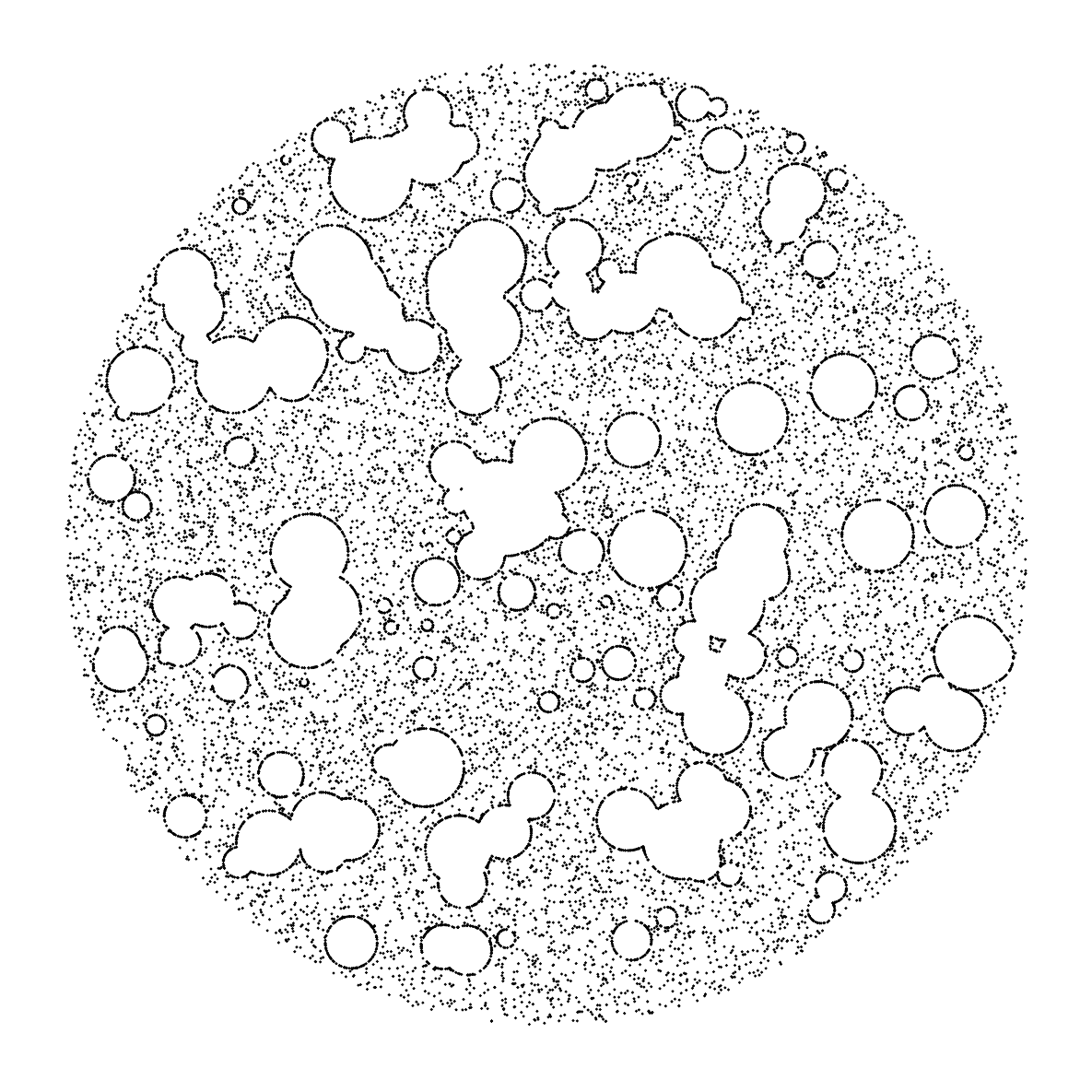}
\includegraphics[width=0.32\linewidth]{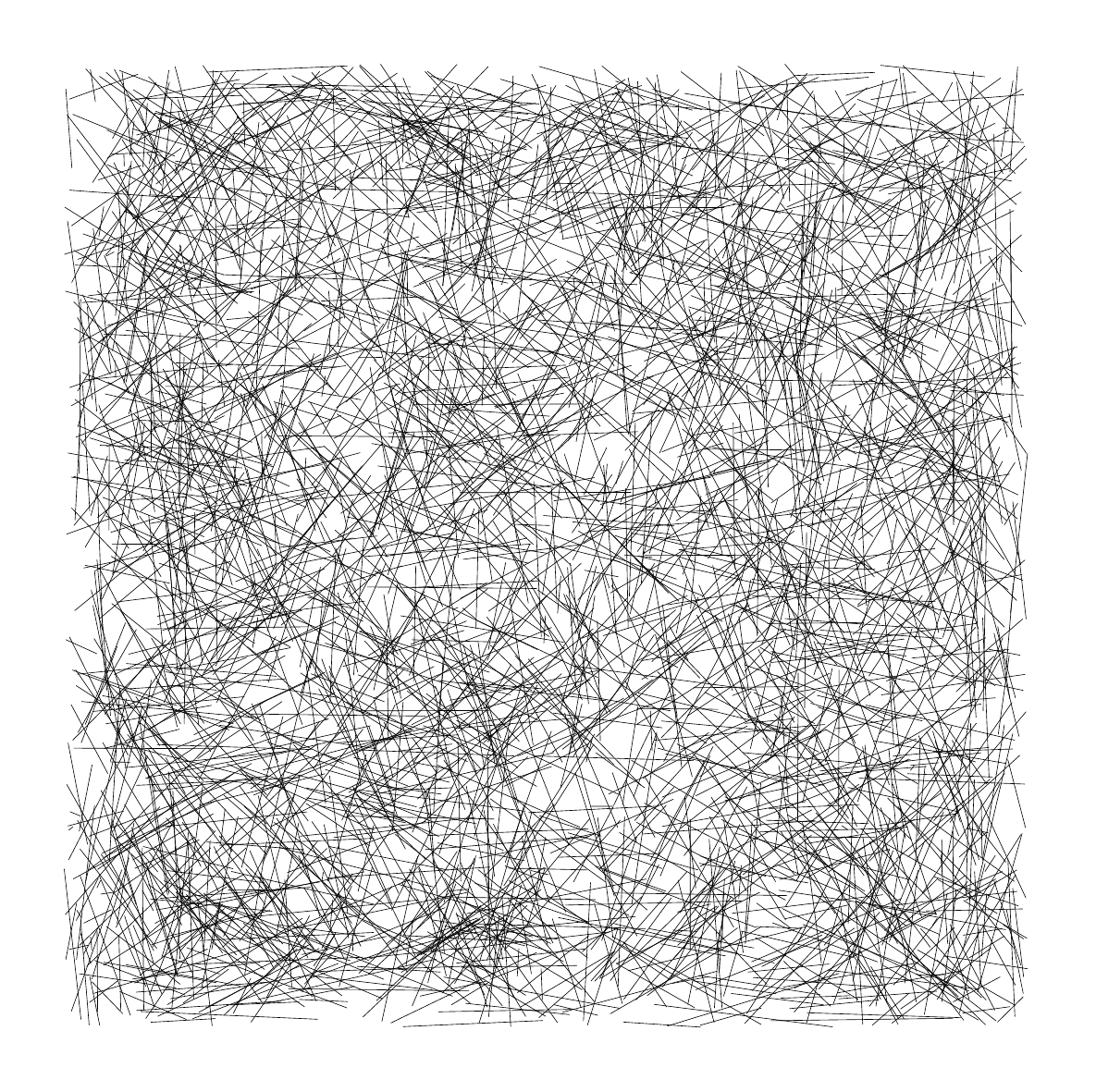}
\includegraphics[width=0.32\linewidth]{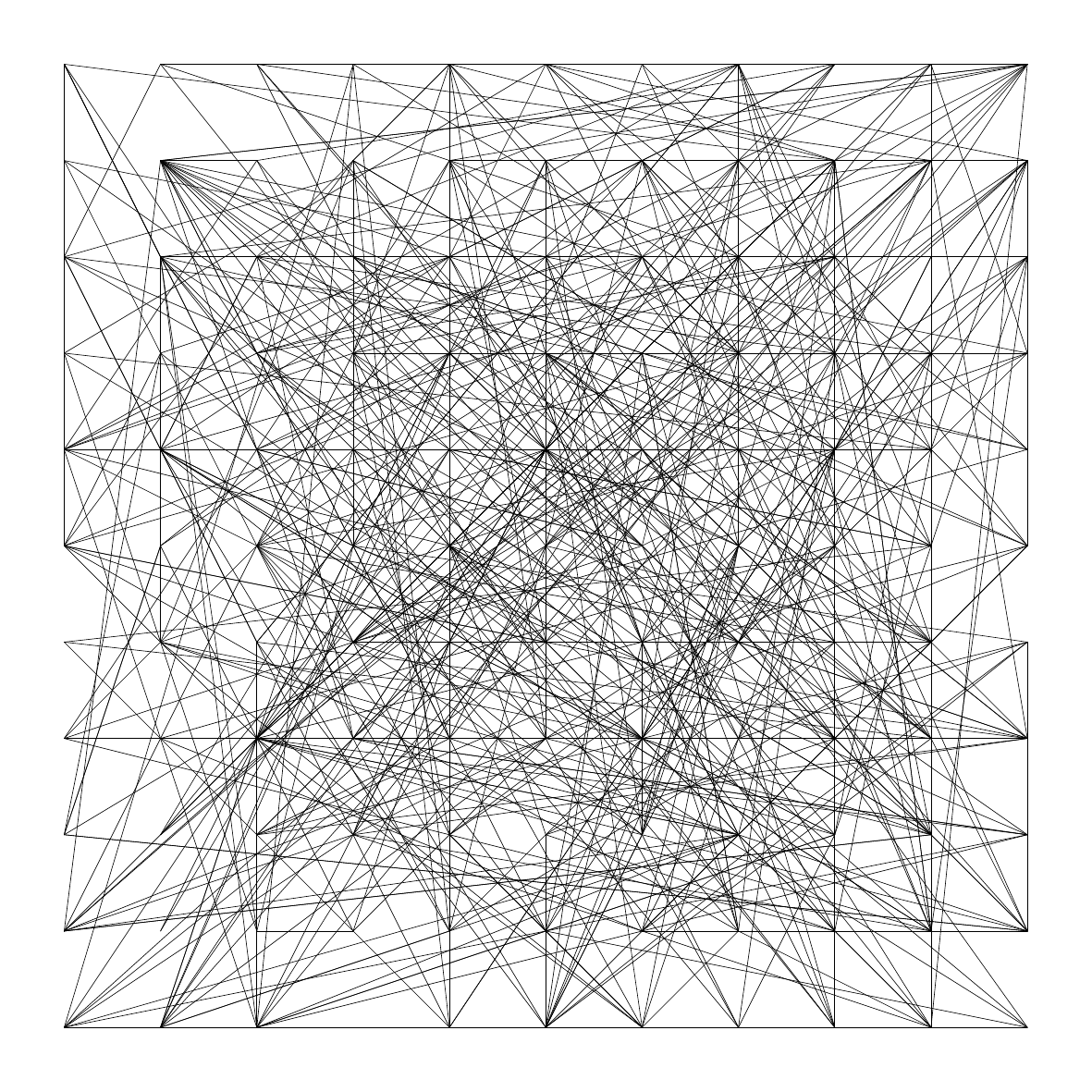}
\caption{Examples from the test data for the delaunay triangulation and intersection points computation. Three types of data are shown: A point cloud with $50\%$ of its points on a union of disks (left), a random collection of short segments (middle) and a random collection of segments with endpoints on a grid (right).}
\label{fig:testdata}
\end{figure}

We compare the three strategies introduced at the beginning of Section~\ref{sec:experiments}. In none of the experiments, a timestamped separation bound computation significantly worsened the performance. So we show only results with a timestamped computation enabled. Instead we show the impact of an exact implementation of \texttt{ceil\_log2} (cf.\ Section~\ref{ssc:switchingerrors}), indicated by an \texttt{x}.

We first compute the delaunay triangulation of a set of $20000$ points, of which between $50\%$ and $100\%$ lie on a union of disks with no points inside (cf.\ Figure~\ref{fig:testdata}).
\begin{figure}[ht]
\begin{tikzpicture}[scale=0.5]
\begin{axis}[
	symbolic x coords={def,lgi,lgd,defx,lgix,lgdx},
	xtick=data,
	ylabel=Time (seconds),
	ymax = 1.1,
    width  = 1.9\linewidth,
    height = 0.8\linewidth,
    bar width = 0.7cm,
    ybar,
    label style={font=\Large},
    tick label style={font=\Large},
    legend style={font=\Large,at={(0.5,0.97)},anchor=north,column sep=5pt},
    legend columns=-1,
    nodes near coords,
    every node near coord/.append style={rotate=90, anchor=west, 
    									 /pgf/number format/precision=2,font=\Large}
]
% delaunay 50 10000
%\addplot[fill=SEQCOL1] coordinates {(def,0.1122) (lgi,0.10008) (lgd,0.10464) (defx,0.21456) (lgix,0.16648) (lgdx,0.17276)};

% delaunay 75 10000
%\addplot[fill=SEQCOL2] coordinates {(def,0.138) (lgi,0.11528) (lgd,0.12356) (defx,0.28508) (lgix,0.21172) (lgdx,0.22612)};

% delaunay 100 10000
%\addplot[fill=SEQCOL3] coordinates {(def,0.1638) (lgi,0.13616) (lgd,0.14764) (defx,0.35344) (lgix,0.25908) (lgdx,0.2708)};

% delaunay 50 20000
\addplot[fill=SEQCOL1] coordinates {(def,0.21676) (lgi,0.18904) (lgd,0.2048) (defx,0.406) (lgix,0.31652) (lgdx,0.33776)};

% delaunay 75 20000
\addplot[fill=SEQCOL2] coordinates {(def,0.2666) (lgi,0.22788) (lgd,0.24932) (defx,0.54236) (lgix,0.40892) (lgdx,0.41964)};

% delaunay 100 20000
\addplot[fill=SEQCOL3] coordinates {(def,0.31424) (lgi,0.26192) (lgd,0.28408) (defx,0.65456) (lgix,0.4732) (lgdx,0.4972)};

\legend{$50\%$,$75\%$,$100\%$}
\end{axis}
\end{tikzpicture}
\caption{Experimental results for the computation of a delaunay triangulation on $20000$ random points with different percentages of them distributed along the boundary of a union of disks and none inside. More degeneracies lead to more performance gain through a logarithmic error bound. An exact computation of \texttt{ceil\_log2} causes an increase in running time in all cases.}
\label{fig:delaunay}
\end{figure}
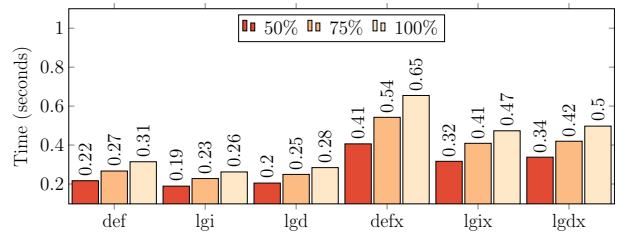

The results are shown in Figure~\ref{fig:delaunay}. The data shows a small performance gain by switching from a direct error representation to a logarithmic one. Using exact transformations between bigfloats and integer types shows to be very expensive in all cases, but even more so in the case of direct error representation. In \texttt{lgi} and \texttt{lgd} errors are not transformed through this method. Still, both experience an increase in running time, since the computation of the magnitude of an \emph{approximation} is affected by this change as well (cf.\ Section~\ref{ssc:switchingerrors}). It can be reasoned that for \texttt{lgi} and \texttt{lgd} the inexact transformation can be used, since the problems from Section~\ref{ssc:switchingerrors} are solved by design. While not fully comparable, the difference between \texttt{defx} and \texttt{lgi} is worth noting.

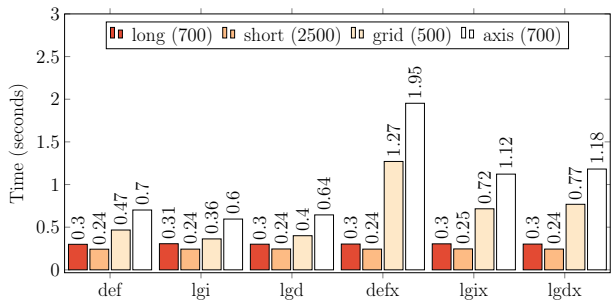
\begin{figure}[ht]

\begin{tikzpicture}[scale=0.5]
\begin{axis}[
	symbolic x coords={def,lgi,lgd,defx,lgix,lgdx},
	xtick=data,
	ylabel=Time (seconds),
	ymax = 3,
    width  = 1.9\linewidth,
    height = \linewidth,
    bar width = 0.5cm,
    ybar,
    label style={font=\Large},
    tick label style={font=\Large},
    legend style={font=\Large,at={(0.5,0.97)},anchor=north,column sep=5pt},
    legend columns=0,
    nodes near coords,
    every node near coord/.append style={rotate=90, anchor=west, 
    									 /pgf/number format/precision=2,font=\Large}
]
% interhom long 700
\addplot[fill=SEQCOL1] coordinates {(def,0.29928) (lgi,0.30612) (lgd,0.3006) (defx,0.30248) (lgix,0.30476) (lgdx,0.30196)};

% interhom short 2500
\addplot[fill=SEQCOL2] coordinates {(def,0.24204) (lgi,0.2428) (lgd,0.24456) (defx,0.24284) (lgix,0.2452) (lgdx,0.24336)};

% interhom grid 500
\addplot[fill=SEQCOL3] coordinates {(def,0.46684) (lgi,0.36256) (lgd,0.4) (defx,1.27048) (lgix,0.71556) (lgdx,0.76832)};

% interhom axis 700
\addplot[fill=white] coordinates {(def,0.70132) (lgi,0.59528) (lgd,0.644) (defx,1.95248) (lgix,1.12204) (lgdx,1.18152)};

\legend{long ($700$),short ($2500$),grid ($500$),axis ($700$)}
\end{axis}
\end{tikzpicture}
\caption{Results for computing the intersection points of segments in various constellations. Neither the form of error representation, nor the exactness of the \texttt{ceil\_log2} operation have an influence on the running time for random long or short segments. If segments are placed on a grid or parallel to the axes, a logarithmic error bound leads to better results. Whether the \texttt{ceil\_log2} operation is performed exactly has significant influence on the performance in these cases.}
\label{fig:intersection}
\end{figure}

In Figure~\ref{fig:intersection} the results for the computation of intersection points on different test data are shown. The test data consists of long or short random segments, segments with endpoints on a grid and segments which are parallel to the axes (cf.\ Figure~\ref{fig:testdata}). The intersection tests are performed with homogenous coordinates. Results for Cartesian coordinates are not shown, but look similar. The same number of segments as in Mörig et al.\ are used in the four scenarios to make the data somewhat comparable.

For long or short random segments, no difference between the three forms of error representation is apparent. Because of the random distribution of the segments, the data sets have almost no degeneracies. Therefore almost all relevant signs can be decided through a floating-point filter, without invoking the accuracy-driven computation in the first place.

For segments on a grid or axis-parallel segments, significant differences are present between different error representations, especially if the transformation between representations is exact. In these cases lots of degeneracies occur, causing the computation to switch to accuracy-driven computation more often.

Summarizing the results, in each geometric experiment the logarithmic representation performed at least as well as the direct representation, while avoiding possible drawbacks due to a faulty error conversion. The more degeneracies occur, the larger is the gain compared to other strategies. Switching to an exact implementation of \texttt{ceil\_log2} is too expensive to be a reasonable alternative for solving the problems resulting from an inexact transformation.

\section{Conclusion}

Choosing a representation model for error bounds in accuracy-driven computation means balancing a tradeoff between the quality of the bound and the efficiency of its computation. Our results suggest that in most cases it is not worth the effort to compute a better error bound to increase the overall efficiency of the number type. A simple logarithmic integer bound outperforms the direct error bound representation as well as the logarithmic floating-point bound in all cases if accompanied by some cautionary mechanisms regarding the separation bound computation.

Mixed forms of error representation suffer from problems arising during the transformation between the representation or, respectively, from the transformation overhead. In conclusion, we suggest using logarithmic integer error bounds by default in exact-decisions number types based on accuracy-driven computation.

\section{Future work}

The additional precision gained through the usage of a logarithmic floating-point representation proved to be not sufficient to compensate for the additional cost associated with its operations in the fixed-precision initialization phase of the algorithm. For accuracy-driven computation it seems promising to combine this approach with a global error propagation strategy as described in Section~\ref{ssc:logdbl}.
To a lesser extent, global strategies may also be used to improve the error bound computed with integer exponents.

\small
\bibliographystyle{abbrv}
\bibliography{lit}
\end{document}